\documentclass[sigconf,natbib=true]{acmart}
\makeatletter
\def\@ACM@checkaffil{
    \if@ACM@countrypresent\else
        \ClassWarningNoLine{\@classname}{No country present for an affiliation}%
    \fi
}
\makeatother
\renewcommand\footnotetextcopyrightpermission[1]{} 

\usepackage{enumitem}

\AtBeginDocument{%
  \providecommand\BibTeX{{%
    \normalfont B\kern-0.5em{\scshape i\kern-0.25em b}\kern-0.8em\TeX}}}

\begin{document}

\title{Attentive Graph-based Text-aware Preference Modeling for Top-N Recommendation}

\author{Ming-Hao Juan}
\affiliation{
  \institution{National Taiwan University}
}
\email{r09922083@csie.ntu.edu.tw}

\author{Pu-Jen Cheng}
\affiliation{
  \institution{National Taiwan University}
}
\email{pjcheng@csie.ntu.edu.tw}

\author{Hui-Neng Hsu}
\affiliation{
  \institution{National Taiwan University}
}
\email{r09922045@csie.ntu.edu.tw}

\author{Pin-Hsin Hsiao}
\affiliation{
  \institution{National Taiwan University}
}
\email{r09944007@csie.ntu.edu.tw}

\renewcommand{\shortauthors}{Juan \emph{et al.}}

\begin{abstract}
Textual data are commonly used as auxiliary information for modeling user preference nowadays. While many prior works utilize user reviews for rating prediction, few focus on top-N recommendation, and even few try to incorporate item textual contents such as title and description. Though delivering promising performance for rating prediction, we empirically find that many review-based models cannot perform comparably well on top-N recommendation. Also, user reviews are not available in some recommendation scenarios, while item textual contents are more prevalent. On the other hand, recent graph convolutional network (GCN) based models demonstrate state-of-the-art performance for top-N recommendation. Thus, in this work, we aim to further improve top-N recommendation by effectively modeling both item textual content and high-order connectivity in user-item graph. We propose a new model named Attentive Graph-based Text-aware Recommendation Model (AGTM). Extensive experiments are provided to justify the rationality and effectiveness of our model design.

\end{abstract}



\maketitle
\pagestyle{plain} 

\section{Introduction}
Collaborative filtering (CF) has been a fundamental method for achieving effective personalized recommendation. A common paradigm for CF is to learn an embedding for every user and item and predict preference based on the learned embeddings~\cite{BPR, HOPRec, NCF, CMN}. Inspired by graph convolutional network (GCN)~\cite{GCN}, recent methods~\cite{GCMC, PinSage, NGCF, LightGCN} utilize similar neighborhood aggregation structures to model the high-order connectivity in user-item interaction graph. Those works demonstrate promising results, showing that modeling high-order connectivity effectively is crucial to top-N recommendation.

Items often come with textual data in recommendation scenarios nowadays.
These textual data include user reviews and item-related contents such as title and description. Many prior works utilize user reviews for rating prediction~\cite{HFT, DeepCoNN, NARRE, Dattn, ANR, DAML, MPCN, NRCA}. Although recent ones deliver promising performance, we empirically find that many cannot perform comparably well on top-N recommendation, which is a more practical way to evaluate a recommendation model in real-world systems, even if we modify them and employ BPR loss~\cite{BPR}. Furthermore, user reviews are not available in some recommendation scenarios, such as one of our experimental datasets: Movielens-TMDB. Only item-related contents are available in this dataset. We believe that item textual content is a more commonly available auxiliary information in various real-world systems than user review, and it provides more specific and accurate information about an item. Therefore, in this work, we focus on utilizing item textual contents to improve the performance of top-N recommendation. Since none of the review-based methods models high-order connectivity in user-item graph, we also aim to integrate it into our method.


To our best knowledge, other than the few works~\cite{JRL, AARM} that focus on review-based top-N recommendation, only one prior work, TPR~\cite{TPR}, dedicates to dealing with this task. TPR treats every word in an item's textual content as a node connecting to the item itself, and the representation learning is conducted on the combined graph. However, due to its design, TPR can only model word-level semantics, ignoring global semantics of a sentence. Also, we believe there is a better way to incorporate textual information without adding nodes to user-item interaction graph.

We propose a new model named Attentive Graph-based Text-aware Recommendation Model (AGTM).
We adopt a framework that consists of two training stages: (1) Textual Content Modeling. We first extract item representations from textual contents. As pretrained Transformer-based language models such as BERT~\cite{BERT} show promising performance on a wide range of downstream tasks, we apply pretrained BERT-based encoder to extract contextualized representations from textual contents. These representations are then condensed into embeddings with a lower dimension by an Autoencoder. 
(2) Item-based Preference Modeling. We then model user preferences based on the extracted item representations. Our goals are to distill useful information from the item representations and to capture high-order connectivity information from user-item graph effectively. To achieve both of our goals, we designed an aspect-aware attentive network to extract initial user representations from items and apply GCN for high-order connectivity modeling.


To summarize, this work makes the following main contributions:
(1) We explore a relatively untouched task and find a new approach toward modeling unstructured textual information and high-order connectivity in user-item graph for top-N recommendation.
(2) We propose a concise aspect-aware attention mechanism to bridge the modeling of item textual content and neighborhood information, effectively integrating item semantics into initial user embeddings.

\section{Methodology}

\begin{figure}[t]
	\centering
	\includegraphics[width=0.48\textwidth]{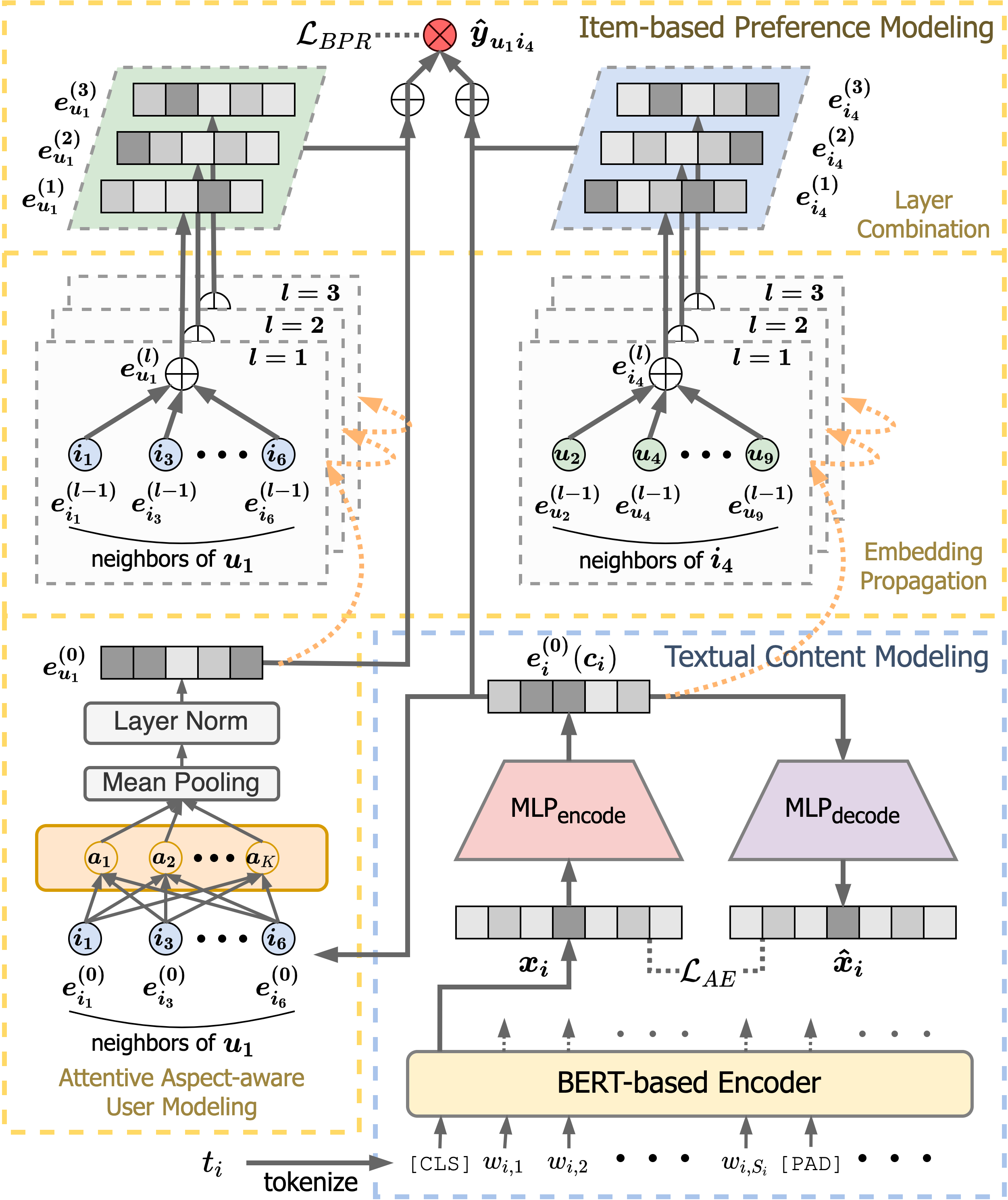}\vspace{-10pt}
	\caption{The model architecture of AGTM. For \textit{item-based preference modeling}, we demonstrate the computation on item $i_4$ and user $u_1$.}
	\label{fig:AGTM}
	\vspace{-12pt}
\end{figure}

We first introduce the problem formulation and then present the proposed Attentive Graph-based Text-aware Recommendation Model (AGTM), as illustrated in Figure~\ref{fig:AGTM}. 

\subsection{Problem Formulation}\label{sec:formulation}
Let $\mathcal{U}$ be the set of users, $\mathcal{I}$ be the set of items, and $\mathcal{T}$ be the set of item textual contents. We denote $\mathcal{R}^+ = \{(u, i)| u \in \mathcal{U}, i \in \mathcal{I}\}$ as the set of observed user-item interactions, where each $(u, i)$ pair represents an interaction between user $u$ and item $i$. Also, for every item $i$, its corresponding textual content is denoted by $t_i \in \mathcal{T}$.

The goal of our model is to learn a function that can predict how likely a user will interact with an unseen item given interaction data $\mathcal{R}^+$ and item textual contents $\mathcal{T}$.

\subsection{Textual Content Modeling}\label{sec:itcm}

\subsubsection{\textbf{Extracting Semantics from Textual Contents}}
We utilize pretrained BERT-based encoder from the Sentence-Transformers (Sentence-BERT~\cite{SBERT}) package\footnote{\url{https://www.sbert.net/index.html}} to extract global semantics from textual contents.
First, we tokenize textual content $t_i$ of item $i$ into BERT-based WordPiece tokens $w_{i,1}, w_{i,2}, ..., w_{i,S_i}$, where $S_i$ is the sequence length of $t_i$. The token sequence prepended with \texttt{[CLS]} is then passed into the BERT-based encoder, which computes a contextualized representation for each token. The output representation of the \texttt{[CLS]} token is the representation of the whole sequence that carries global semantics.


\subsubsection{\textbf{Condensing Content Representations}}\label{sec:condense}
The output embedding size $d_o$ of most pretrained BERT-based encoders is at least 384, much larger than the commonly used 32$\sim$128 in CF models. Training with such a large embedding size will consume too much time and memory. Therefore, we must condense the output BERT representation to a smaller size.

Here we adopt Autoencoder~\cite{AE}. For an item $i$ and its textual content $t_i$, let the representation of \texttt{[CLS]} token be $\textbf{x}_i \in \mathbb{R}^{d_o}$. We use a 2-layer MLP (Multi-Layer Perceptron) as encoder, and a symmetric 2-layer MLP as decoder. The encoding and decoding of Autoencoder can be abstracted as
\begin{equation}
\begin{aligned}
    \textbf{c}_i = \text{MLP}_{\text{encode}}(\textbf{x}_i), \quad
    \hat{\textbf{x}}_i = \text{MLP}_{\text{decode}}(\textbf{c}_i),
\end{aligned}
\end{equation}
where $\textbf{c}_i \in \mathbb{R}^{d}$ is the condensed embedding for further training in the next stage, $d$ is the reduced embedding size, and $\hat{\textbf{x}}_i$ is the reconstructed representation of \texttt{[CLS]} token.

We train the Autoencoder first by all $\textbf{x}_i$ from $\mathcal{T}$ before continuing to the next stage. (Training details are given in Section~\ref{sec:training}.) 

\subsection{Item-Based Preference Modeling}\label{sec:upm}
After extracting item semantics from textual contents, we move on to model user preferences.


\subsubsection{\textbf{Attentive Aspect-aware User Modeling}}
As we now have item initial embedding $\textbf{e}_i^{(0)} = \textbf{c}_i$, we have to first obtain user initial embedding $\textbf{e}_u^{(0)}$ for representation learning.
Intuitively, a user's preference can be described by their interacted items. As items already possess rich content information, we assume that a user's representation can first be obtained from neighboring items. Thus, we designed an attentive network to extract information from item representations. We perform masked attention on the item nodes, only attending to items that have interaction with a user. The attention coefficient of a neighboring item $i$ of user $u$ is computed as
\begin{equation}\label{eq:attn}
    \alpha_u^i = \text{Softmax}(\textbf{a}^\top\textbf{e}_i^{(0)}) =  \frac{\text{exp}(\textbf{a}^\top\textbf{e}_i^{(0)})}{\sum_{j \in \mathcal{N}_u} \text{exp}(\textbf{a}^\top\textbf{e}_j^{(0)})},
\end{equation}
where $\textbf{a} \in \mathbb{R}^d$ is a shared weight vector, $\mathcal{N}_u$ denotes the set of items that user $u$ interacted with.

Usually, a user looks at multiple aspects of an item to decide whether or not to interact with it, so we employ multi-head attention to extract different aspects from items. Specifically, there are $K$ independent attention mechanisms computing Equation~\ref{eq:attn}, each with a different weight vector $\textbf{a}_k$ ($\textbf{a}$ in Equation~\ref{eq:attn}), and then we employ mean pooling to gather the obtained $K$ representations, followed by layer normalization~\cite{LayerNorm}, written as
\begin{equation}
    \textbf{e}_u^{(0)} = \text{LayerNorm}(\frac{1}{K} \sum_{k=1}^{K} \sum_{i \in \mathcal{N}_u} \alpha_{u,k}^i \textbf{e}_i^{(0)}).
\end{equation}

Two major differences between our method and many review-based methods are that we do not apply any linear transformation in the attention network~\cite{NARRE, Dattn, ANR, DAML, MPCN, NRCA, AARM} and that we remove the use of randomly initialized ID embeddings~\cite{NARRE, NRCA}. We believe that they bring no benefit but negatively increase the difficulty of training.


\subsubsection{\textbf{Embedding Propagation and Layer Combination}}
We then stack multiple graph convolutional layers to capture high-order connectivity information. To do it effectively, we adopt similar weighted sum aggregation as \textit{Light Graph Convolution} in LightGCN~\cite{LightGCN}. The propagation rule is defined as
\begin{equation}\label{eq:emb-prop}
\begin{aligned}
    \textbf{e}_u^{(l+1)} = \sum_{i \in \mathcal{N}_u} \frac{1}{\sqrt{|\mathcal{N}_u||\mathcal{N}_i|}} \textbf{e}_i^{(l)}, \quad
    \textbf{e}_i^{(l+1)} = \sum_{u \in \mathcal{N}_i} \frac{1}{\sqrt{|\mathcal{N}_u||\mathcal{N}_i|}} \textbf{e}_u^{(l)},
\end{aligned}
\end{equation}
where $\textbf{e}_u^{(l)}$ and $\textbf{e}_i^{(l)}$ respectively denote the refined embedding of user $u$ and item $i$ after $l$ layers of propagation, and $\mathcal{N}_u$ ($\mathcal{N}_i$) denotes the set of items (users) that connect to user $u$ (item $i$) in the user-item interaction graph.


After computing $L$ layers of propagation by Equation~\ref{eq:emb-prop}, we want the final representation to include different information captured in different layers, making it more comprehensive. Therefore, we combine the embeddings obtained at each layer to form the final representation by mean pooling, written as
\begin{equation}
    \textbf{e}_u = \frac{1}{L}\sum_{l=0}^L \textbf{e}_u^{(l)}, \quad \textbf{e}_i = \frac{1}{L}\sum_{l=0}^L \textbf{e}_i^{(l)},
\end{equation}
where $\textbf{e}_u$ and $\textbf{e}_i$ denote the final representations of $u$ and $i$.


\subsection{Model Prediction and Training}\label{sec:training}
For model prediction, we apply inner product to estimate a user's preference to a target item, written as
\begin{equation}
    \hat y_{ui} = \textbf{e}_u \cdot \textbf{e}_i,
\end{equation}
which is used as the ranking score for recommendation generation.

As for training AGTM, there are two stages. First, we train the Autoencoder in \textit{textual content modeling}. We apply Mean Squared Error (MSE) as the loss function, defined as
\begin{equation}
    \mathcal{L}_{\text{AE}} = \sum_{i \in \mathcal{I}}(\hat{\textbf{x}}_i - \textbf{x}_i)^2 + \lambda_1 \Vert \Theta_{\text{AE}} \Vert_2^2,
\end{equation}
where $\Theta_{\text{AE}}$ denotes all trainable parameters in the Autoencoder, and $\lambda_1$ is the $L_2$ regularization coefficient. Note that we do not further fine-tune the BERT-based encoder since semantically similar sentences are already close in vector space after Sentence-BERT pre-training.

For the next stage, \textit{item-based preference modeling}, we apply the \textit{Bayesian Personalized Ranking} (BPR) loss~\cite{BPR}, 
written as
\begin{equation}
    \mathcal{L}_{\text{BPR}} = - \sum_{u \in \mathcal{U}} \sum_{(i,j) \in \mathcal{O}_u} \ln \sigma (\hat y_{ui} - \hat y_{uj}) + \lambda_2 \Vert \Theta_{\text{IPM}} \Vert_2^2,
\end{equation}
where $\mathcal{O}_u = \{(i,j) | i \in \mathcal{N}_u, j \notin \mathcal{N}_u\}$ is the pair-wise training data for user $u$, $\sigma(\cdot)$ is the Sigmoid function, $\lambda_2$ is the $L_2$ regularization coefficient, and $\Theta_{\text{IPM}}$ only includes initial item embeddings $\textbf{e}_i^{(0)}$ and $K$ weight vectors $\textbf{a}_k$.

\section{Experiments}


\subsection{Experimental Settings}

\subsubsection{\textbf{Dataset Description}}

\begin{table}[t]
\caption{Dataset statistics.}\label{tab:exp-dataset}
\vspace{-10pt}
\resizebox{0.5\textwidth}{!}{
\hskip-10pt
\begin{tabular}{l|rrrrr}
\hline
& \multicolumn{1}{c}{\small\textbf{\# User}} & \multicolumn{1}{c}{\small\textbf{\# Item}} & \multicolumn{1}{c}{\small\textbf{\# Interaction}} & \multicolumn{1}{c}{\small\textbf{\# Item w/ text}} & \multicolumn{1}{c}{\small\textbf{Density}} \\ \hline\hline
Amazon-Grocery  &  $12,464$ &  $6,644$ &   $205,739$ &  $6,639$ & $0.00248$ \\ \hline
Amazon-CDs      &  $21,450$ & $18,398$ &   $527,503$ & $16,486$ & $0.00134$ \\ \hline
MovieLens-TMDB  &   $6,040$ &  $3,260$ &   $998,539$ &  $3,240$ & $0.05071$ \\ \hline
\end{tabular}
}
\vspace{-10px}
\end{table}

We conduct experiments on three real-world datasets: Amazon-Grocery, Amazon-CDs (CDs and Vinyl), and MovieLens-TMDB. The first two datasets are from Amazon-Review (2018) dataset collection\footnote{\url{https://nijianmo.github.io/amazon/index.html}}~\cite{AmazonReview}. We take \textit{title} and \textit{description} in the metadata file and concatenate them as the textual content of an item. 
The third dataset is a combined dataset based on MovieLens-1M~\cite{MovieLens}. We download movie overview (introduction) from TMDB\footnote{\url{https://www.themoviedb.org/}} by its official API and concatenate it after movie title as the textual content of a movie.

We apply the 10-core setting for each dataset, i.e., discarding users and items with less than ten interactions. The statistics of datasets are summarized in Table~\ref{tab:exp-dataset}. We randomly split the user-item interactions into three subsets: training, validation, and testing sets, each with a proportion of 75\%, 5\%, and 20\%, respectively. 

\subsubsection{\textbf{Evaluation Metrics}}
We apply two widely-used evaluation protocols~\cite{TPR, NGCF, LightGCN}: Recall@20 and NDCG@20.
Also, we apply the all-ranking strategy~\cite{SampledMetrics, LightGCN, NGCF}. Specifically, all items that have no interaction with a user are seen as negative, and the interacted items in the testing set are seen as positive.

\subsubsection{\textbf{Compared Methods}}
\begin{itemize}[leftmargin=*]
    \item Pure CF methods: \textbf{MF-BPR}~\cite{BPR}, \textbf{LightGCN}~\cite{LightGCN}. LightGCN is the state-of-the-art GCN-based recommendation model.
    \item Review-based methods: \textbf{DeepCoNN-BPR}~\cite{DeepCoNN}, \textbf{NARRE-BPR}~\cite{NARRE}, \textbf{AARM}~\cite{AARM}. Directly adapting rating prediction models (DeepCoNN and NARRE) to top-N recommendation, which is to rank items by rating predictions, leads to very poor performance (both Recall@20 and NDCG@20 $\leq$ 0.02). Hence, we remove the prediction layers of these models, and take the final user and item embeddings to compute BPR loss, using the exact method that we train our AGTM model. On the other hand, AARM is the state-of-the-art model among the few works that focus on review-based top-N recommendation. Also, since there is no user review available in MovieLens-TMDB, we skip those experiments.
    \item \textbf{TPR}~\cite{TPR}: It is the state-of-the-art recommendation model that incorporates item textual content.
\end{itemize}

\subsubsection{\textbf{Hyper-parameter Settings}}
We implement our AGTM model in PyTorch. For BERT-based encoder, we choose the best performing pretrained model in Sentence-BERT\footnote{\url{https://www.sbert.net/docs/pretrained_models.html}} to date: \textit{all-mpnet-base-v2}, with output embedding size $d_o = 768$. The initial embedding size $d$ is fixed to 64 for all models. We optimize all models except TPR\footnote{We directly use the code from authors: \url{https://github.com/cnclabs/codes.tpr.rec}} with AdamW~\cite{AdamW} optimizer, and train with batch size 1024.
We conduct grid search to find the optimal setting for each method. The learning rate is searched in \{$10^{-3}$, $3\times 10^{-3}$\},  the coefficient of $L_2$ regularization $\lambda_2$ is searched in \{$10^{-4}$, $10^{-3}$, $10^{-2}$, $10^{-1}$\}, and the number of aspects $K$ for AGTM is searched in \{2, 4, 8\}. 
Furthermore, we conduct validation for every 10 epochs, and training will stop if NDCG@20 on the validation set does not improve for 100 epochs. 

\subsection{Performance Comparison}

\begin{table}[t]
\caption{Overall performance comparison.}\label{tab:exp-overall}
\vspace{-10px}
\resizebox{0.5\textwidth}{!}{
\hskip-10pt
\begin{tabular}{l|c c|c c|c c}
\hline
& \multicolumn{2}{c|}{\small\textbf{Amazon-Grocery}} & \multicolumn{2}{c|}{\small\textbf{Amazon-CDs}} & \multicolumn{2}{c}{\small\textbf{MovieLens-TMDB}} \\ 
                      &           Recall &             NDCG &           Recall &             NDCG &           Recall &             NDCG \\ \hline\hline
MF-BPR                &           0.1189 &           0.0890 &           0.1384 &           0.0919 &           0.2212 &           0.3271 \\ 
LightGCN-3$^\dagger$  &           0.1494 &\underline{0.1187}&\underline{0.1672}&\underline{0.1153}&\underline{0.2494}&\underline{0.3751}\\ \hline
TPR                   &\underline{0.1686}&           0.1172 &           0.1627 &           0.1076 &           0.2417 &           0.3699 \\ \hline
DeepCoNN-BPR          &           0.0882 &           0.0545 &           0.0739 &           0.0465 &                - &                - \\
NARRE-BPR             &           0.1192 &           0.0851 &           0.0965 &           0.0608 &                - &                - \\
AARM                  &           0.1445 &           0.1012 &           0.1484 &           0.0996 &                - &                - \\ \hline
AGTM-3$^\dagger$      &   \textbf{0.1747}$^\ast$&   \textbf{0.1312}$^\ast$&   \textbf{0.1756}$^\ast$&   \textbf{0.1209}$^\ast$&   \textbf{0.2590}$^\ast$&   \textbf{0.3930}$^\ast$\\ \hline\hline
Improvement           &          3.62 \% &         10.53 \% &          5.02 \% &          4.86 \% &          3.85 \% &          4.77 \% \\
$p$-value             &          1.97e-4 &          6.54e-6 &          4.36e-4 &          6.35e-8 &          1.01e-5 &          1.37e-7 \\ \hline
\end{tabular}
}
\vskip 1pt
\baselineskip 1pt
\raggedright{\footnotesize{$^\dagger$-3 denotes 3 GCN layers.}}

\raggedright{\footnotesize{$^\ast$ Statistically significant ($p<0.01$) in $t$-test, relative to the best competing method (underlined).}}
\vspace{-5pt}
\end{table}

The results are reported in Table~\ref{tab:exp-overall}. We find that:
\begin{itemize}[leftmargin=*]
    \item Even though we employ BPR loss for DeepCoNN and NARRE, they still yield poor performance. We believe that this is because the designs of both models are too heavy and burdensome for top-N recommendation, which increase the difficulty of training. \cite{LightGCN} has a similar finding which suggests that performing multiple feature transformations brings no benefit to the model effectiveness for CF tasks. On the other hand, AARM delivers better performance, but it still performs worse than LightGCN, which does not incorporate any auxiliary information. It highlights the importance of modeling high-order connectivity.
    
    \item TPR only performs better than LightGCN on Amazon-Grocery in terms of Recall@20. The reason might be that item textual contents in Amazon-Grocery generally contain more information, helping the model capture extra CF signals. However, when the information is not rich enough, the model cannot benefit from the extra information. We believe it is because of the complicated graph TPR used for representation learning and the lack of global semantics when modeling textual contents.
    \item AGTM consistently yields the best performance on all datasets, demonstrating the effectiveness of our model design. The improvements are all statistically significant as $p$-value < 0.05. 
\end{itemize}

\subsection{Ablation Studies}

\subsubsection{\textbf{Effect of Main Components and Two-stage Approach}}\label{sec:exp-component}

\begin{table}[t]
\caption{Effect of main components and two-stage approach.}\label{tab:exp-item-user}
\vspace{-10pt}
\resizebox{0.5\textwidth}{!}{
\hskip-10pt
\begin{tabular}{l|c c|c c|c c}
\hline
& \multicolumn{2}{c|}{\small\textbf{Amazon-Grocery}} & \multicolumn{2}{c|}{\small\textbf{Amazon-CDs}} & \multicolumn{2}{c}{\small\textbf{MovieLens-TMDB}} \\ 
                              &         Recall &           NDCG &         Recall &           NDCG &         Recall &           NDCG \\ \hline\hline
AGTM-3                        & \textbf{0.1747}& \textbf{0.1312}& \textbf{0.1756}& \textbf{0.1209}& \textbf{0.2590}& \textbf{0.3930}\\ \hline
AGTM-3\textsubscript{w/o TCM} &         0.1527 &         0.1180 &         0.1682 &         0.1161 &         0.2547 &         0.3898 \\
AGTM-3\textsubscript{w/o AAUM} &         0.1511 &         0.1179 &         0.1668 &         0.1132 &         0.2490 &         0.3744 \\ 
AGTM-3\textsubscript{w/o IPM} &         0.1681 &         0.1287 &         0.1544 &         0.1012 &         0.2481 &         0.3826 \\ 
AGTM-3\textsubscript{w/o AE}  &         0.1598 &         0.1132 &         0.1276 &         0.0844 &         0.2307 &         0.3544 \\ \hline
\end{tabular}
}
\vspace{-5pt}
\end{table}

The following variants are constructed: (1) AGTM-3\textsubscript{w/o TCM}: We randomly initialize item ID embedding $\textbf{e}_i^{(0)}$, removing information from \textit{textual content modeling} (TCM).
(2) AGTM-3\textsubscript{w/o AAUM}: We remove attentive aspect-aware user modeling (AAUM) in \textit{item-based preference modeling}, and add randomly initialized user ID embedding $\textbf{e}_u^{(0)}$ for training.
(3) AGTM-3\textsubscript{w/o IPM}: We further remove the whole \textit{item-based preference modeling} (IPM), and directly train $\textbf{e}_i^{(0)}$ and the randomly initialized $\textbf{e}_u^{(0)}$.
(4) AGTM-3\textsubscript{w/o AE}: We replace the Autoencoder with a linear transformation layer, making $\textbf{e}_i^{(0)} = \textbf{W} \textbf{x}_i$, so $\textbf{x}_i$ and $\textbf{W}$ are trained together with \textit{item-based preference modeling}, i.e., training altogether in one stage.

We summarize the results in Table~\ref{tab:exp-item-user}. We can find that both \textit{textual content modeling} (TCM) and \textit{item-based preference modeling} (IPM) are effective and indispensable. Removing any one of them leads to a substantial performance drop. Also, our attentive aspect-aware user modeling (AAUM) design eliminates the need for training user ID embeddings and still improves the performance, demonstrating its effectiveness. Jointly analyzing Tables~\ref{tab:exp-overall} and \ref{tab:exp-item-user}, we can notice that AGTM-3\textsubscript{w/o TCM} even slightly outperforms LightGCN-3. Both models do not utilize any auxiliary information. Therefore, our attentive aspect-aware user modeling design is effective even without textual information from items.

For the two-stage training approach, as AGTM-3\textsubscript{w/o AE} underperforms AGTM-3 with a large margin, it demonstrates the rationality of our two-stage training, which is to condense item embeddings with Autoencoder first.

\subsubsection{\textbf{Effect of Number of GCN Layers}}\label{sec:exp-n-layer}

\begin{table}[t]
\caption{Effect of number of GCN layers.}\label{tab:exp-n-layer}
\vspace{-10pt}
\resizebox{0.48\textwidth}{!}{
\begin{tabular}{l|c c|c c|c c}
\hline
& \multicolumn{2}{c|}{\small\textbf{Amazon-Grocery}} & \multicolumn{2}{c|}{\small\textbf{Amazon-CDs}} & \multicolumn{2}{c}{\small\textbf{MovieLens-TMDB}} \\ 
       &         Recall &           NDCG &         Recall &           NDCG &         Recall &           NDCG \\ \hline\hline
AGTM-1 &         0.1495 &         0.1079 &         0.1418 &         0.0975 &         0.2542 &         0.3875 \\ 
AGTM-2 &         0.1730 &         0.1284 &         0.1697 &         0.1167 &         0.2589 &         0.3926 \\
AGTM-3 &         0.1747 & \textbf{0.1312}&         0.1756 &         0.1209 & \textbf{0.2590}&         0.3930 \\
AGTM-4 & \textbf{0.1748}&         0.1299 & \textbf{0.1788}& \textbf{0.1220}&         0.2578 & \textbf{0.3933}\\ \hline
\end{tabular}
}
\vspace{-12pt}
\end{table}

We then analyze to what extent AGTM can benefit from multiple embedding propagation layers. We conduct experiments with the number of GCN layers in \{1, 2, 3, 4\}.

Table~\ref{tab:exp-n-layer} summarizes the experimental results. We find that increasing the layers from 1 to 2 improves the performance significantly. The benefits diminish after that. Using 3 GCN layers leads to satisfactory performance in most cases. Furthermore, jointly analyzing Tables~\ref{tab:exp-overall} and \ref{tab:exp-n-layer}, we find that AGTM-2 already outperforms all competing methods except LightGCN-4 on all datasets. This again verifies the effectiveness of the AGTM design.

\section{Conclusion}
In this work, we find a new approach to model unstructured textual information for top-N recommendation. We propose AGTM, which jointly models item textual content and high-order connectivity in user-item interaction graph. The key of AGTM is the use of BERT-based encoder to extract contextualized representation from text and the novel attentive aspect-aware user modeling followed by GCN to distill item information and model user preference effectively. Extensive experiments on three real-world datasets demonstrate the effectiveness and rationality of our model design.



\bibliographystyle{ACM-Reference-Format}
\bibliography{bibfile}

\end{document}